\documentclass[doublecol]{epl2} 

\usepackage{comment}
\usepackage{amsmath}
\usepackage{color}

\title{Network flow of mobile agents enhances the evolution of cooperation}
\shorttitle{} 

\author{G. Ichinose\inst{1} \and Y. Satotani\inst{2} \and T. Nagatani\inst{3}}
\shortauthor{G. Ichinose \etal}

\institute{                    
  \inst{1} Department of Mathematical and Systems Engineering, Shizuoka University, Hamamatsu 432-8561, Japan\\
  \inst{2} Department of Information Technology, Okayama University, Okayama 700-8530 Japan\\
  \inst{3} Department of Mechanical Engineering, Shizuoka University, Hamamatsu 432-8561, Japan
}
\pacs{87.23.Kg}{Dynamics of evolution}
\pacs{89.65.-s}{Social and economic systems}
\pacs{87.23.Cc}{Population dynamics and ecological pattern formation}

\abstract{
We study the effect of contingent movement on the persistence of cooperation on complex networks with empty nodes.
Each agent plays Prisoner's Dilemma game with its neighbors and then it either updates the strategy depending on the payoff difference with neighbors or it moves to another empty node if not satisfied with its own payoff.
If no neighboring node is empty, each agent stays at the same site.
By extensive evolutionary simulations, we show that the medium density of agents enhances cooperation where the network flow of mobile agents is also medium.
Moreover, if the movements of agents are more frequent than the strategy updating, cooperation is further promoted.
In scale-free networks, the optimal density for cooperation is lower than other networks because agents get stuck at hubs.
Our study suggests that keeping a smooth network flow is significant for the persistence of cooperation in ever-changing societies.
}

\begin{document}

\maketitle

\section{Introduction}
One important biological and social issue is the persistence of cooperation.
Cooperation is disadvantageous compared to defection because such behavior results in some costs to oneself by giving benefits to others.
In contrast, defection receives the benefits without paying any cost.
Therefore, cooperation cannot be an evolutionarily stable strategy for a noniterative game in a well-mixed population \cite{nowak06evolutionaryDynamicsBOOK, MaynardSmith1973, TaylorJonker1978, MaynardSmithBOOK1982, HofbauerBOOK1998}.

In such a situation, network reciprocity can account for the persistence of cooperation \cite{NowakMay1992, SantosPacheco2005} (See also some reviews \cite{PercSzolnoki2010, Perc_etal2013}).
Some pioneering studies used spatial Prisoner's Dilemma (PD) game to consider the effect of network reciprocity on the evolution of cooperation where agents' locations are fixed.
In some co-evolutionary models of network topology and strategy \cite{Tanimoto2007, Tanimoto2009, ZimmermannEguiluz2005, Pacheco_etal2006, SuzukiKatoArita2008}, agents can change interacting partners by severing one link and building a new link. This behavior promotes cooperation when the rewiring links create cooperative groups. Even in this case, agents' locations are still fixed.
However, in reality, biological organisms usually interact with others at locations that are not fixed.
Rather, they interact with others by moving if spaces are available.
In the first attempt of the evolution of cooperation with mobility, spaces are not explicitly considered \cite{EnquistLeimar1993}.
Then, various models which explicitly assume two dimensional space (discrete or continuous) with empty sites have been developed \cite{DugatkinWilson1991, Vainstein_etal2007, Jiang_etal2010, Yang_etal2010, Aktipis2004, HelbingYu2008, Helbing2009, HelbingYu2009, Ichinose_etal2013, Antonioni_etal2014, SuzukiKimura2011, Meloni_etal2009, Ichinose_etal2013b, Vainstein_etal2014, AmorFort2011, Cardillo_etal2012, Sicardi_etal2009, TomassiniAntonioni2015}.

In those studies, a square lattice is usually used as the discrete interaction environment. For continuous environments, two dimensional continuous space is used.
In both cases, the environments are homogeneous.
This may not be appropriate to describe real interactions because they could be biased due to the heterogeneity of environments.
For example, in human society, people do not interact at every place. When people are walking on a road, they just pass by without interaction.
In the case where a person yields to another on a road, there are some interactions between them, but still weak. 
Interactions take place where people gather such as schools, local groups, and offices etc.
To realize the heterogeneous environments, we use complex networks for the interaction space and reveal the effect of heterogeneity.

Another important aspect is the type of movements.
First, unconditional movements have been studied for the evolution of cooperation \cite{Vainstein_etal2007, Sicardi_etal2009, Vainstein_etal2014, Antonioni_etal2014, SuzukiKimura2011, Meloni_etal2009, Cardillo_etal2012}. Unconditional movements means that agents move to randomly selected empty sites with some probability.
In this case, frequent movements prohibit cooperation because cooperative clusters, which are needed for cooperation to evolve, are destroyed, therefore moderate mobility facilitates cooperation \cite{Vainstein_etal2007, Meloni_etal2009}.
Cooperation is significantly enhanced when contingent movements are introduced \cite{Jiang_etal2010, Yang_etal2010, Aktipis2004, HelbingYu2008, Helbing2009, HelbingYu2009, Ichinose_etal2013, Ichinose_etal2013b, AmorFort2011, Sicardi_etal2009, TomassiniAntonioni2015}.
Here we assume contingent movements because it only requires minimum intelligence such as detecting the condition of the present location, which allows us to apply the mechanism to a broad range of biological organisms.

In the present paper, we investigate the contingent movements of agents on complex networks with empty nodes.
Note that, in our model, the underlying network topology never changes. However, their actual interactions (who plays the PD with whom)  change because agents move around on empty nodes.
We show that, when the density of agents are medium, cooperation is significantly enhanced.
The optimal density decreases in the case of scale-free networks because agents get stuck at hubs.

\section{Model}
We developed an agent-based model in which $n$ agents play the PD game, update their strategy, and change the location by movements on complex networks.
We compared three different networks; Random Regular Graph (RRG), Lattice, and Barab\'asi-Albert scale-free networks (BA) \cite{BarabasiAlbert1999}.
RRG and Lattice can be classified as homogeneous networks because all nodes have the same degree, while BA is different from those homogeneous networks and can be classified as a heterogeneous network.
In BA, a small number of nodes, called hubs, connect with a substantial number of links while most other nodes connect with a few others.
Each network size is represented by $N$ with the average degree $k$ and the density of agents is $\rho$. Thus, $n$ is given by $n=N \rho$.

Initially, $n$ agents are randomly placed on the nodes in a network.
Each agent can take one of two strategies in the PD game: Cooperation ($C$) or Defection ($D$).  At first, half of the agents are allocated $C$ and the other half  allocated $D$.

In each time step, one agent, denoted $i$, is randomly picked up.
Agent $i$ plays the PD game with $k_i$ neighbors and obtains the payoff. The payoff is divided by $k_i$ and the average payoff is denoted by $\pi_i$.
The neighbors also play the PD game with their own neighbors and obtain the payoff.
The neighbors' payoffs are also averaged.
In each game, both individuals obtain payoff $R$ for mutual cooperation or $P$ for mutual defection.
If one selects cooperation while the other does defection, the former obtains the sucker's payoff $S$ while the latter obtains the highest payoff $T$, the temptation to defect.
The relationship of the four payoffs is usually $T > R > P > S$ in the PD games.
For simplicity, we use $T>1, R=1, P=0, S=1-T$. Thus $T$ is the only parameter for PD.
This form of the PD game is called the donor-recipient (or donation) game where two parameters $b$ (benefit) and $c$ (cost) for cooperation are assumed \cite{Nowak2006}. Our parametrization is further simplified because of only one parameter, $T$, which is realized by letting $T=b>1$ and $c=b-1$ from the donor-recipient game.
For more details, pairwise games including PD can be generalized by considering $D_g$ (gamble-intending; GID) and $D_r$ (risk-averting; RAD) dilemmas \cite{TanimotoSagara2007, Wang_etal2015, Tanimoto2009}.
In each time step, either movement or strategy updating always takes place.
With probability $\tau_m$, the movement phase is selected.  In the opposite case, i.e., with probability  $\tau_s=1-\tau_m$, the strategy updating phase is selected.
We can consider $\tau_m$ and $\tau_s$ as the parameters that adjust time-scale of evolution between movement (topological change of interaction) and strategy updating.
Pacheco et al. revealed that cooperation is enhanced when the topological change takes place faster than strategy updating \cite{Pacheco_etal2006}.

When the movement phase is selected, agent $i$ moves to another neighboring empty node with probability $p_{m, i}=\frac{T-\pi_i}{T-S}$. It becomes 1 when the average payoff $\pi_{m, i}$ is equal to $S$. The probability linearly decreases as $\pi_{m, i}$ increases. Finally, it becomes 0 when the average payoff $\pi_{m, i}$ is equal to $T$.
As $S=1-T$, $p_{m, i}$ can be rewritten as $p_{m, i}=\frac{T-\pi_i}{2T-1}$. Agent $i$ does not move if no neighboring node is empty.
This type of movement can be considered contingent movement \cite{Jiang_etal2010, Yang_etal2010, Aktipis2004, HelbingYu2008, Helbing2009, HelbingYu2009, Ichinose_etal2013, Ichinose_etal2013b, AmorFort2011, Sicardi_etal2009} because an agent tends to move to another node if the current node is unfavorable in the sense that the current location gives a lower payoff. In those models, on the other hand, agents do not move when they are satisfied with the payoff in the current location.
Instead, when the strategy updating phase is selected, we use the pairwise comparison rule \cite{TraulsenNowakPacheco2006, TraulsenPachecoNowak2007, SzaboToke1998}.
In short, agent $i$ randomly selects one of the neighbors $j$ and agent $i$ imitates $j$'s strategy with the probability 
\begin{equation}
p_{s, i} = [1+\mathrm{e}^{-\beta (\pi_j -\pi_i)}]^{-1},
\end{equation}
where $\beta \geq 0$ controls the intensity of selection. For $\beta=0$, there is no selection pressure, meaning that evolutionary dynamics proceeds by random drift. As $\beta$ becomes larger, the tendency that strategies with higher payoffs will be imitated increases.
We set $\beta=0.1$.
This falls into the framework of evolutionary games because the strategy with higher fitness will be imitated more often.
These movement and strategy updating phases (either one is selected in each time step) are repeated $G$ generations.
One generation is defined as $n$ time steps where agents are activated (movement or strategy) once on average.
We use $g$ for specifying generation where $0 \leq g \leq G$.

\section{Results}
To check the evolution of cooperation, we calculated the fraction of cooperation in the last 10\% of generations in each simulation.
Because the time scale for the strategy updating $\tau_s$ is different in each setting, we set $G=10^6/(n \tau_s)$ in order to keep the number of strategy updating the same in the different settings.
We conducted 50 independent simulation runs and the average fraction of cooperation was used for describing each data point in the result graphs.
The network size for RRG and BA is $N=5000$ while Lattice is $N=71^2=5041$. The average degree in all networks is $k=4$.
The other three parameters; $T, \tau_m (=1-\tau_s)$, and $\rho$ are swept as the main experimental parameters. In the results, two of them are changed while the other one is fixed.

\begin{figure*}[t]
\begin{center}
\includegraphics[width=\textwidth]{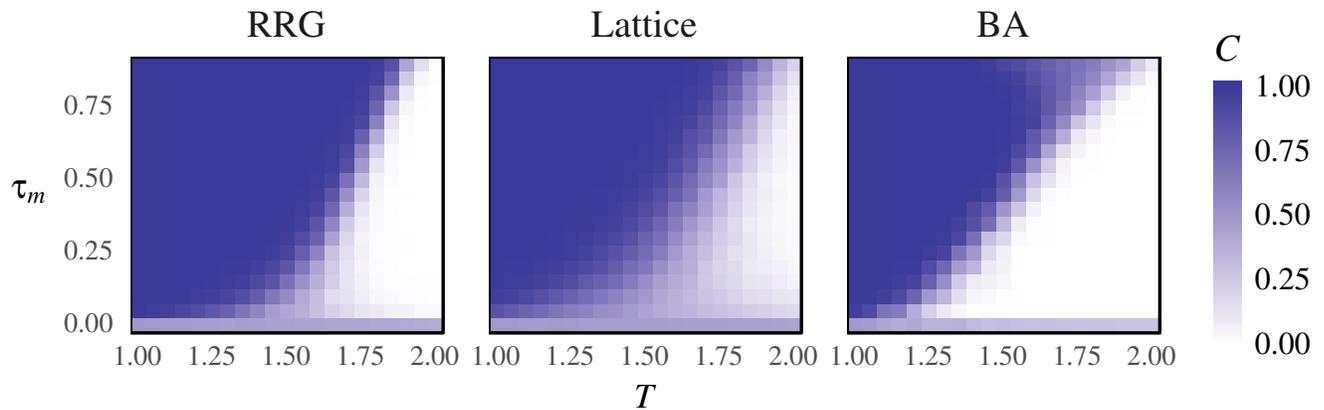}
\caption{Fraction of cooperation on $T - \tau_m$ space. $\rho=0.3$ is fixed.}
\label{T-tau}
\end{center}
\end{figure*}

We first show the fraction of cooperation on $T - \tau_m$ space (Fig.~\ref{T-tau}).
In all networks, cooperation decreases as $T$ increases. This is simply because larger $T$ gives benefits to defectors.
The effect of $\tau_m$ on cooperation is also simple.
Because agents use contingent movements, there is a greater chance for cooperators to make clusters before they are exploited by defectors when $\tau_m$ is large.
In contrast, if $\tau_m$ is low, cooperators become defectors more often because cooperators are weak in the case that cooperative clusters are not generated.
In Lattice networks, the dynamics are slightly different with RRG and BA networks where the middle $\tau_m$ is optimal for cooperation when $T$ is high.
There may be many local cooperative clusters in Lattice networks. In that case, if $\tau_m$ is too high, those local clusters tend to collapse due to the high frequency of movements which has harmful effects on cooperation.

\begin{figure*}[t]
\begin{center}
\includegraphics[width=\textwidth]{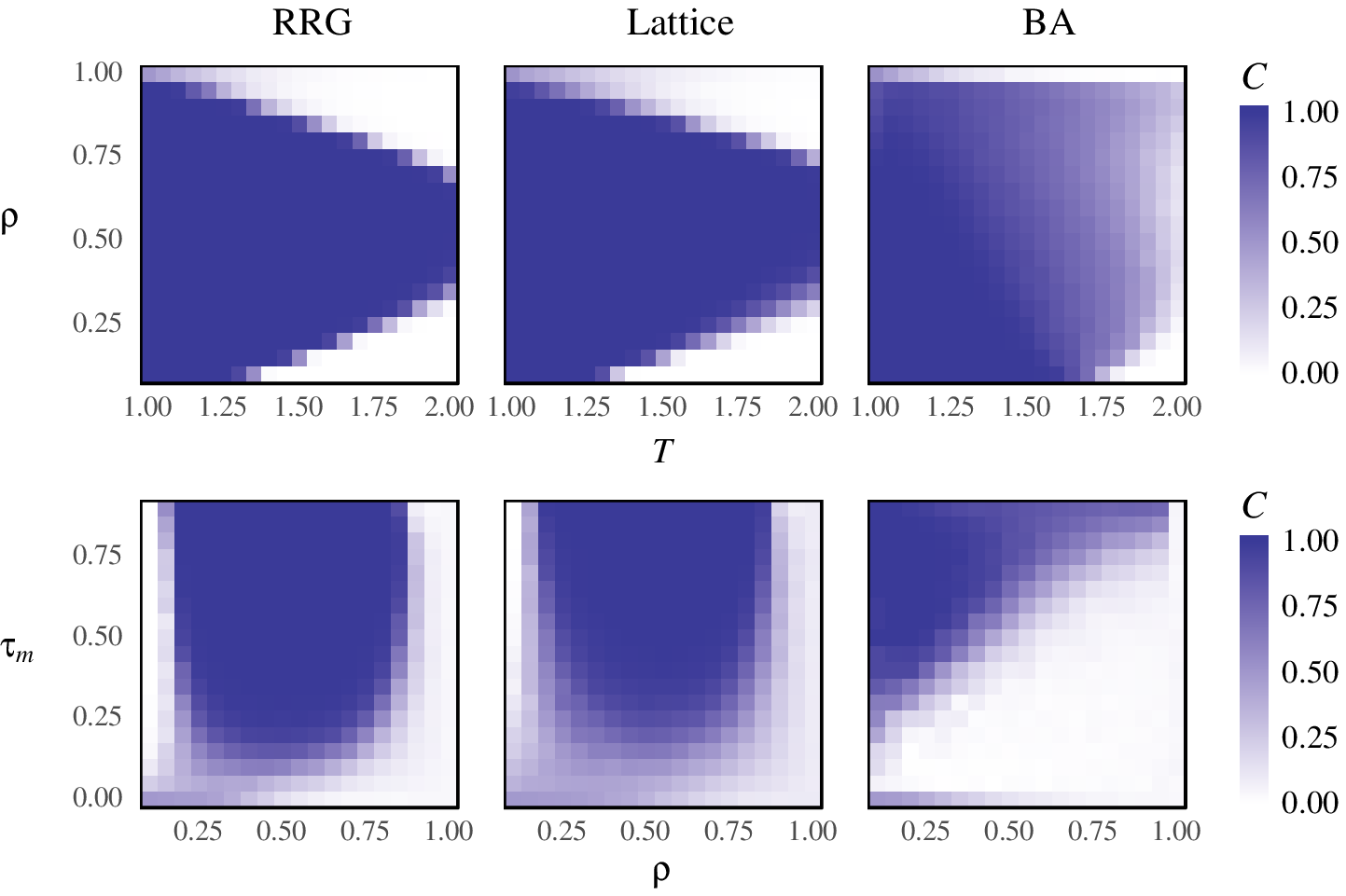}
\caption{Top: Fraction of cooperation on $T - \rho$ space. $\tau_m=0.9$ is fixed. Bottom: Fraction of cooperation on $\rho - \tau_m$ space. $T=1.5$ is fixed.}
\label{T-rho_rho-tau}
\end{center}
\end{figure*}

\begin{figure*}[t]
\begin{center}
\includegraphics[width=\textwidth]{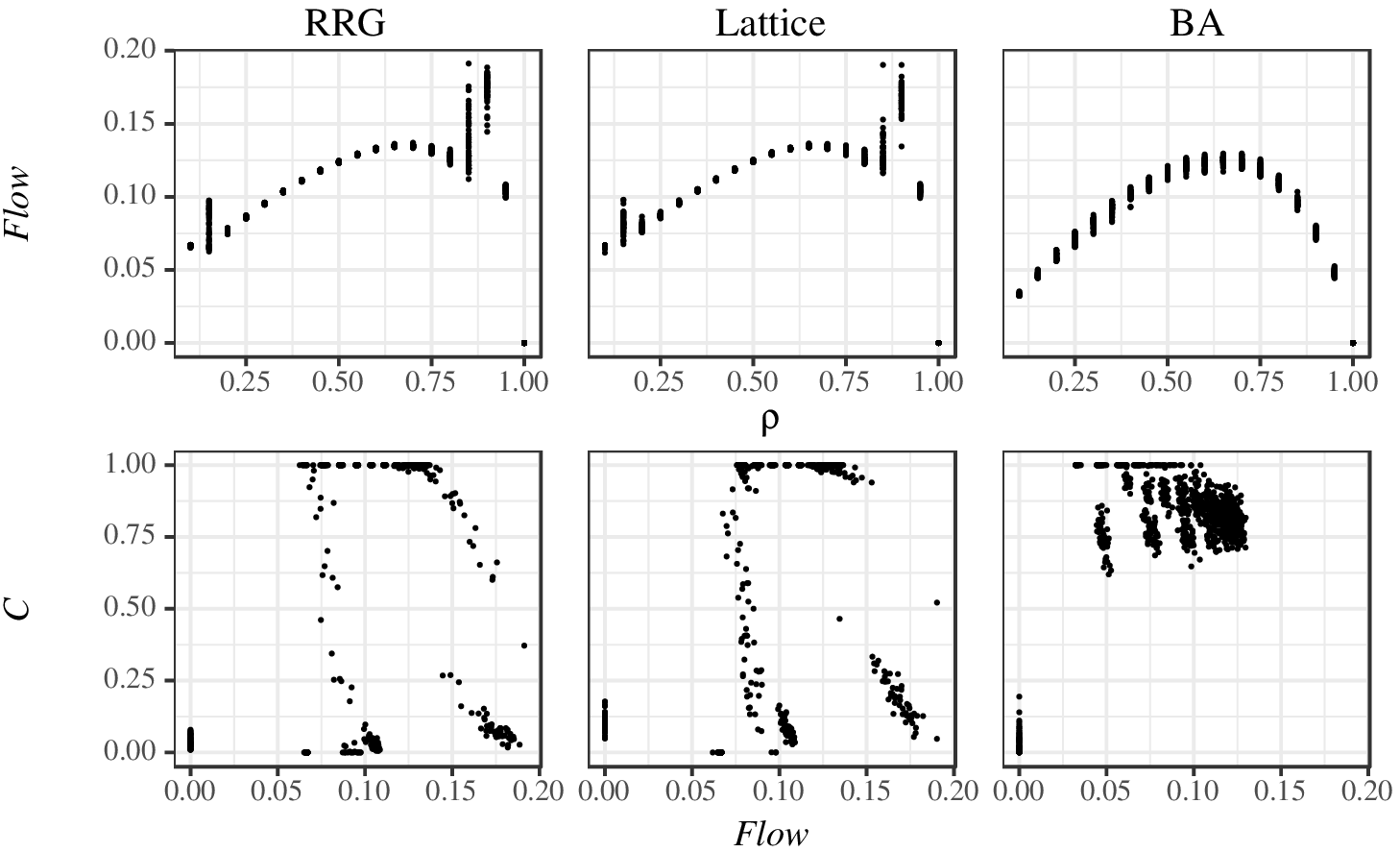}
\caption{Top: Relationship between $\rho$ and Flow. Flow is defined as the value of $\rho$ multiplied by the average rate of movement in one time step. Bottom: Fraction of cooperation as a function of Flow. $T=1.5$ and $\tau_m=0.9$ are fixed. In both cases, one data point is dotted per one simulation instead of averaging 50 simulations.}
\label{flow-and-foc}
\end{center}
\end{figure*}

We next focus on the effect of $\rho$ on cooperation where $\tau_m =0.9$ is fixed (Top in Fig.~\ref{T-rho_rho-tau}).
Interestingly, the maximum cooperation level is obtained in the intermediate value of $\rho$ (around 0.5) in RRG and Lattice networks.
The tendency that the high frequency of movement ($0.5 \leq \tau_m$) and the medium value of density ($0.1 \leq \rho \leq 0.8$) in RRG and Lattice networks promote cooperation the most is also verified in $\rho-\tau_m$ space with $T=1.5$ (Bottom in Fig.~\ref{T-rho_rho-tau}).
To reveal the reason, we conducted additional experiments.
In each simulation of the additional experiments, we calculated the number of average times that agents actually moved.
Then, we defined a new metric ``Flow'' which was obtained as the value of $\rho$ multiplied by the average rate of movement in one time step.
Top in Fig.~\ref{flow-and-foc} shows the relationship between $\rho$ and flow.
In every network, the maximum flow is obtained in the middle of $\rho$ values.
This is because if the density is too high, almost all space is occupied by someone in the neighboring spaces.
In this case, only a few agents can move, resulting in the drastic decrease of the number of movement times.
As a result, the maximum flow is realized in the medium $\rho$.
This is well known in traffic flow where traffic jams occur once the density of cars exceeds the critical limit \cite{ChowdhurySantenSchadschneider2000, Helbing2001, Nagatani2002, Kerner2004}.
Bottom in Fig.~\ref{flow-and-foc} shows the relationship between flow and fraction of cooperation.
We obtain the highest cooperation level when the flow is not too large (between 0.10 and 0.15 in RRG and Lattice networks and between 0.05 and 0.125 in BA networks).
If flow is low, contingent movement does not work well because the network is sparse.
In the opposite case where flow is too high, contingent movement also does not work well because high flow destroys the clusters of cooperation, making a well-mixed situation.
Thus, the intermediate flow most promotes cooperation. We show that network flow is the key for the evolution of cooperation.

\begin{figure*}[t]
\begin{center}
\includegraphics[width=\textwidth]{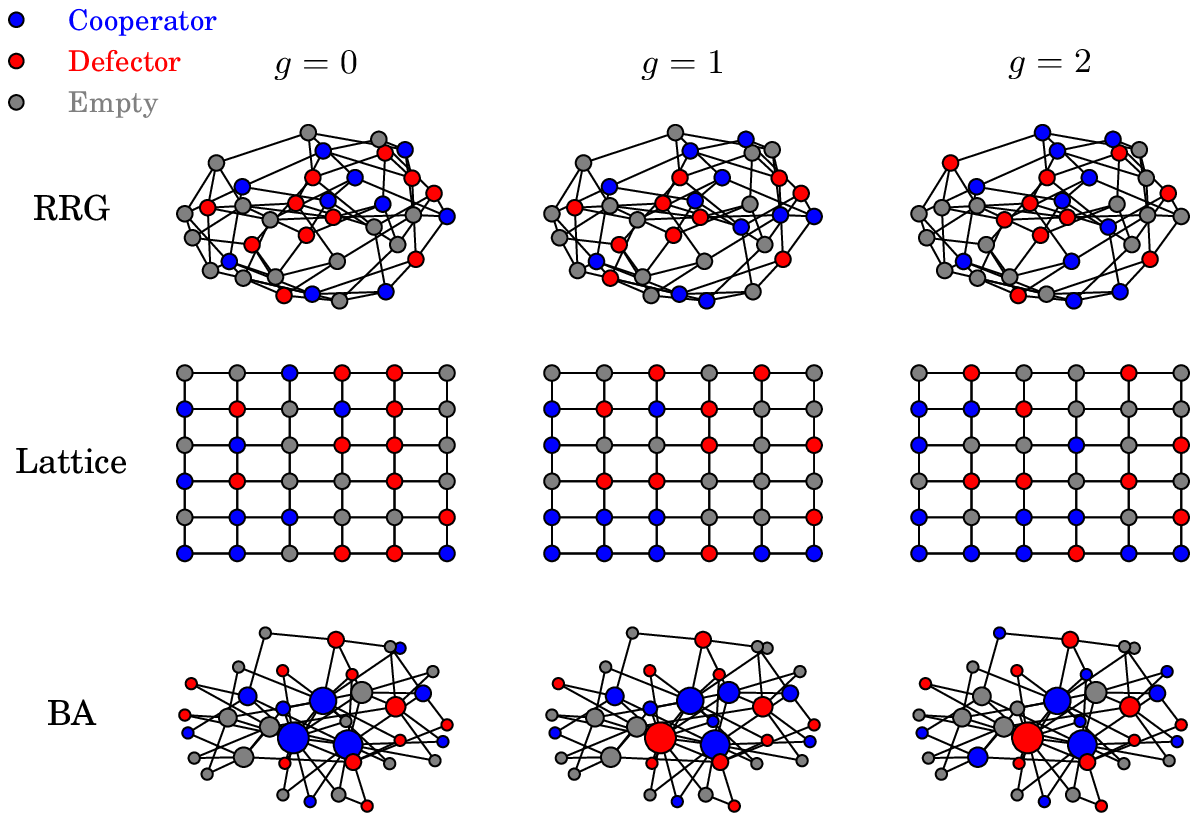}
\caption{Actual dynamics of movement and cooperation on the three networks at initial generations. For visualization, we used quite a small population $N=36$ and $\rho=0.583$ but the dynamics are the same with the simulations used in the results. In the BA network, the degrees of each node are different and the size of each node reflects its degree.
Agents tend to get stuck in the BA network compared to the other networks because the hubs create the bottleneck of network flow.}
\label{net-by-time}
\end{center}
\end{figure*}

The optimal cooperation level in BA networks is obtained lower $\rho$ (Top in Fig.~\ref{T-rho_rho-tau} BA) than the one in RRG and Lattice networks, and so we consider the reason.
Figure \ref{net-by-time} shows the actual dynamics of movement and cooperation on networks at initial generations although each network size is greatly reduced for the sake of visualization ($N=36$).
In BA networks, the locations of agents almost do not change as time goes by ($g=0$ to $g=2$). This is because hubs can be the bottleneck of movements.
Because almost all (peripheral) nodes are only connected to the hubs, agents have to pass through the hubs when they move to other nodes.
Therefore, once hubs are occupied by someone, agents in peripheral nodes cannot move and get stuck in the current node.
Note that this is not the case for agents on hubs.
We investigated the number of times that agents on the largest hub moved.
We set the situation the same as Fig.\ref{net-by-time} and counted the number of times that agents on the hub moved from $g=0$ to 1000.
The results were mean 332.8 times with SD of 23.7 when all the nodes became $C$ and mean 463.9 times with SD of 22.1 when all the nodes became $D$. We ran one hundred simulations for each setting.
The former is less compared to the latter because agents tend not to move when they are satisfied.
Thus, agents on the hub are frequently replaced because there are many empty nodes around the hub.
 See also the supplemental movie which shows what happened to the hub.
Due to this situation, it is difficult for agents on peripheral nodes to move in BA networks compared to the cases in RRG and Lattice networks even when the same $\rho$ is given.
This is why the flow on BA networks is optimal in lower values of $\rho$ compared to RRG and Lattice.

\begin{figure*}[t]
\begin{center}
\includegraphics[width=\textwidth]{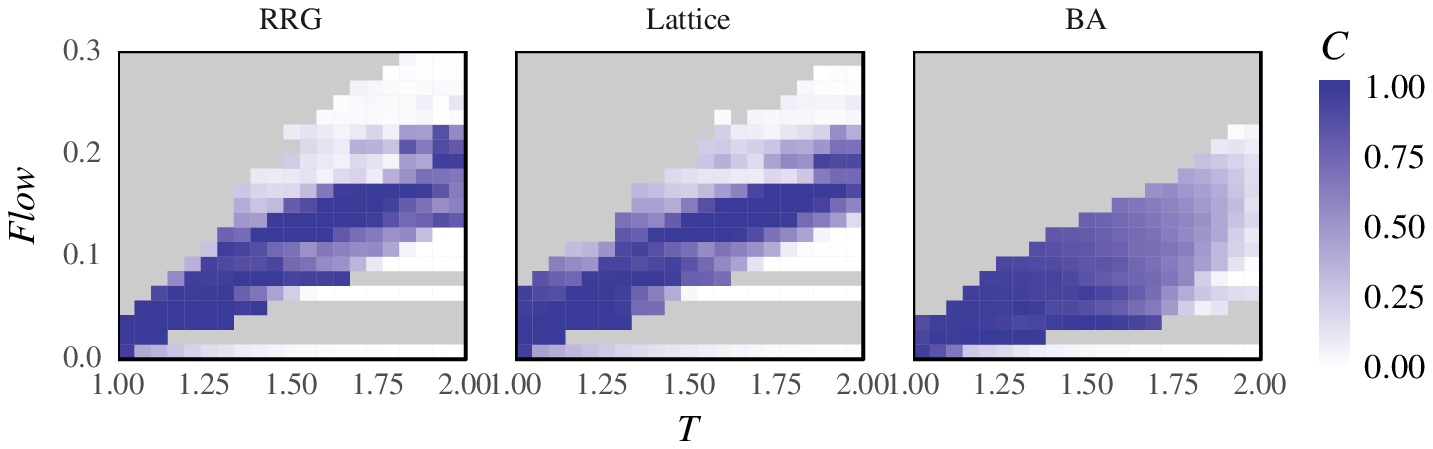}
\caption{Top: Fraction of cooperation on $T - Flow$ space. $\tau_m=0.9$ is fixed. The gray color denotes that  the flow is not realized in those settings.}
\label{foc-by-temp-and-flow}
\end{center}
\end{figure*}

We finally focus on how the optimal flow for cooperation is affected by $T$ (Fig.~\ref{foc-by-temp-and-flow}).
In any networks, if $T$ is low, only the low level flow is realized because almost all agents are satisfied with their payoffs and do not move.
As $T$ becomes larger, flow increases. In this case, cooperation is most promoted when flow is medium, which allows cooperators to make clusters easier.

\section{Conclusion}
In this paper, we studied the effect of network flow on the evolution of cooperation by the means of agent-based evolutionary simulations.
We used two classes of networks: homogeneous (RRG and Lattice) and heterogeneous (BA).
We revealed that, in the presence of contingent movement, the medium density of agents enhances cooperation where the network flow of mobile agents is also medium.
The optimal density for BA networks is relatively lower than RRG and Lattice networks because hubs create the bottleneck of network flow.
Moreover, if the movements of agents are more frequent than the strategy updating, cooperation is further promoted.
We first showed that the flow of the network significantly affects cooperation on networks.

\acknowledgments
G.I. acknowledges the support by Hayao Nakayama Foundation For Science \& Technology \& Culture.

\footnotesize

\end{document}